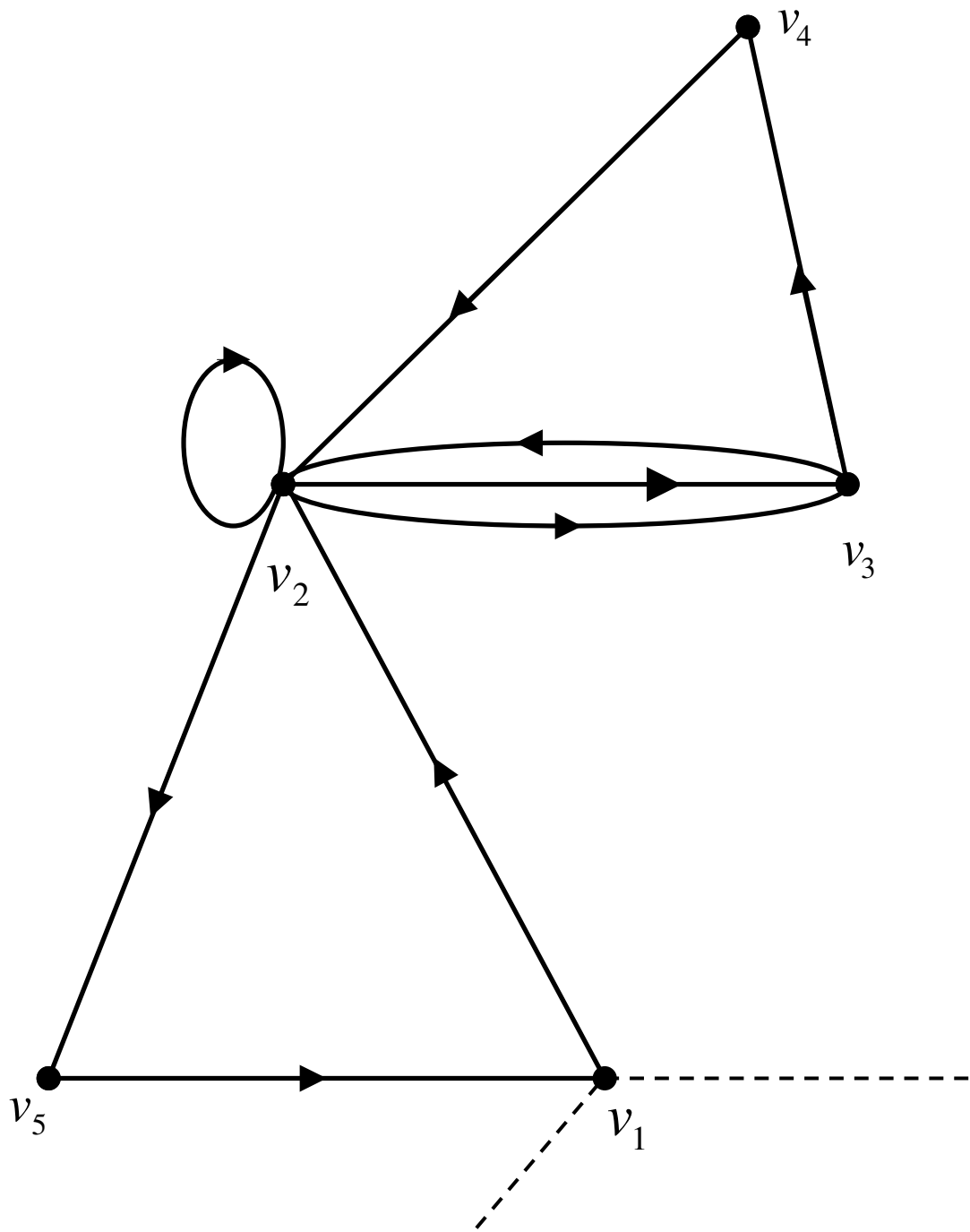

Fig. 1

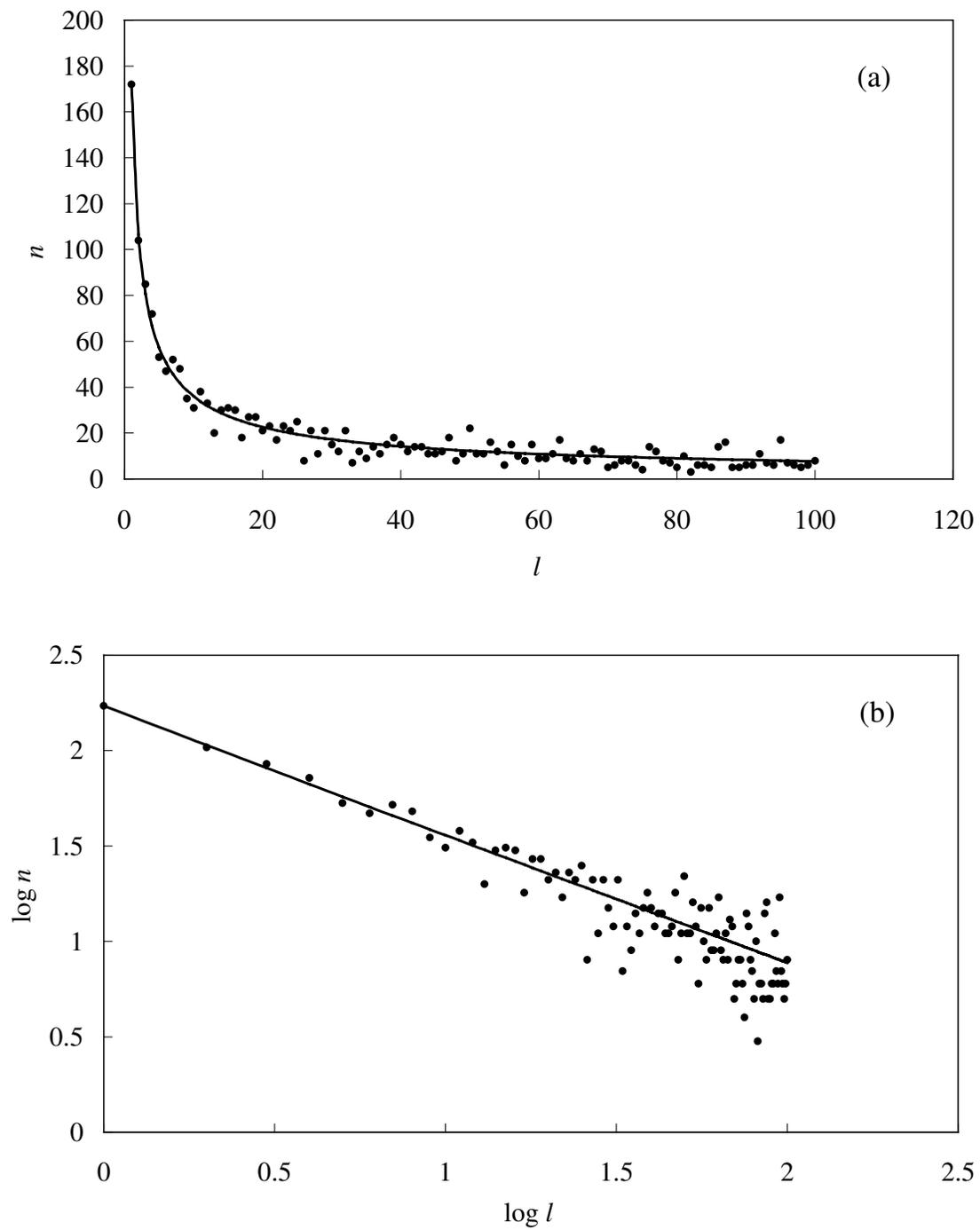

Fig. 2

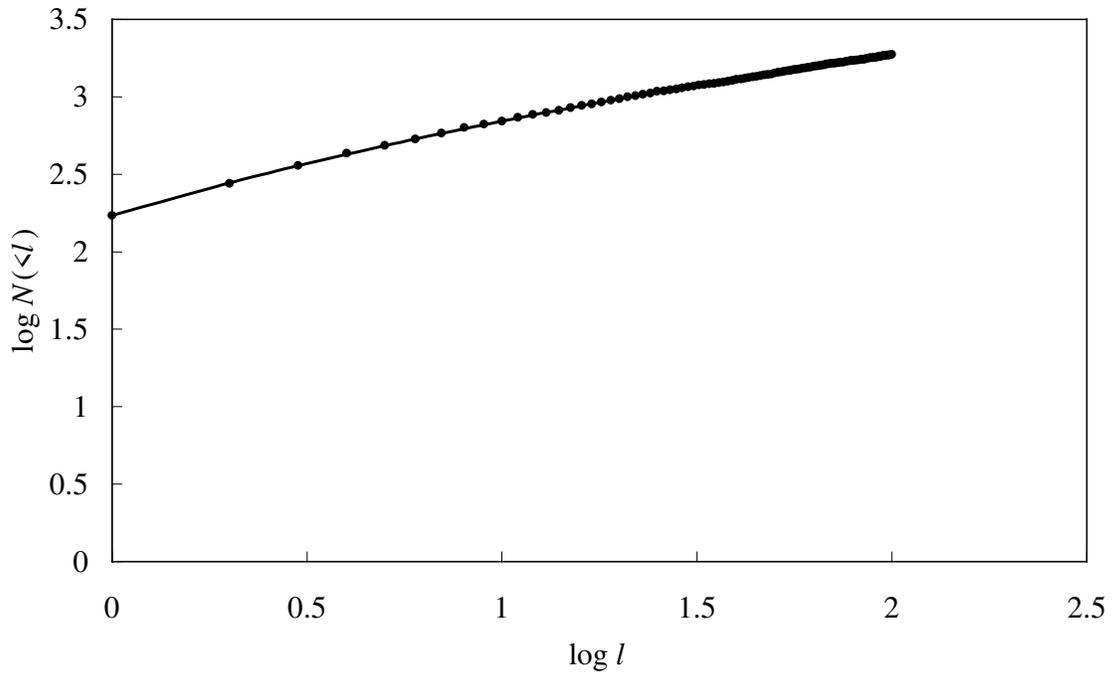

Fig. 3

# Scale-invariant statistics of the degrees of separation in directed earthquake network

S. Abe[1, a] and N. Suzuki[2, b]

[1]Institute of Physics, University of Tsukuba, Ibaraki 305-8571, Japan
[2]College of Science and Technology, Nihon University, Chiba 274-8501, Japan

**Abstract.**    Discovery of a new scale-free law of earthquake phenomenon is reported. It is relevant to the structural and dynamical properties of the earthquake network proposed in a recent work [S. Abe and N. Suzuki, Europhys. Lett. **65**, 581 (2004)]. The seismic data taken in southern California are mapped to an evolving *directed* network. It is found that statistics of the degrees of separation between two vertices is characterized by the Zipf-Mandelbrot distribution. This shows how a given earthquake can influence a large number of subsequent earthquakes and describes spatio-temporal criticality/complexity of seismicity in a novel manner.

**PACS.** 05.65.+b, 91.30.-f, 89.75.Da

___________________________________

[a] e-mail: suabe@sf6.so-net.ne.jp
[b] e-mail: suzu@phys.ge.cst.nihon-u.ac.jp

Earthquake phenomenon has been of continuous interest for researchers in view of science of complexity [1-5]. In particular, understanding of spatio-temporal correlation between earthquakes is of extreme importance in geophysics. For example, there is a work [6], which reports that a major earthquake can trigger aftershocks that are more than 1000 km far from the earthquake. The aftershocks follow the Omori law [7,8], which describes the slow power-law decay of their temporal pattern. These facts suggest that "event correlation" in seismicity may be long-ranged both spatially and temporally. In this respect, we have recently analyzed the seismic data taken in southern California and Japan, and have found that both the distance between two successive earthquakes [9] and the corresponding waiting time (termed "calm time") [10] follow the definite statistical laws described by nonextensive statistical mechanics [11,12], which is expected to be a unified theoretical approach to complex systems. In addition, we have also found [13] that the aging phenomenon and its associated scaling property exist in the nonstationary seismic time series termed the Omori regime in which the aftershocks obey the Omori law, whereas, in the stationary regime outside the Omori regime, no aging can be observed. This result, combined with the slow relaxation (i.e., the power-law nature of the Omori law) and the quenched disorder of the stress distribution during the series of fast aftershocks, suggests that the mechanism governing aftershocks may be of glassy dynamics.

Recently, the concept of complex networks has also been attracting great interest as a novel approach to complex systems. This stream was initiated by the works of Watts and Strogatz on small-world networks [14] and of Barabási and Albert on scale-free networks [15]. The primary purpose of research in this area is to study the topological and dynamical properties of complex systems described as random graphs [16,17]. Quite recently, the concept of complex network has been introduced in seismology in Refs. [18,19]. The earthquake network is constructed as follows. The geographical region under consideration (southern California, in the present case) is divided into a lot of small cubic cells. A cell is regarded as a vertex if earthquakes with any values of magnitude occurred therein. If two successive earthquakes occur in different cells, the corresponding two vertices are linked by an arc (edge), whereas if two successive events occur in an identical cell, then a loop is attached to the vertex. In this way, the fault-fault interactions governed by unknown microscopic dynamics are replaced by the arcs and loops. This procedure enables us to map the seismic data to an evolving directed complex network. There is a unique parameter in this construction, which is the cell size. Since there are no *a priori* rules to determine the cell size, it is important to examine its effects on the structure of the earthquake network. This point was carefully examined in Refs. [18,19], recently. It was found that the earthquake network has some remarkable properties in its topology, which essentially remain unaltered by the change of the cell size. It is a scale-free network [18], and if directedness of the network is ignored, it behaves as a small-world network [19] with small values of the degree of separation between two vertices as well as large values of the clustering coefficient. These findings highlight the complex-system aspect of seismicity in a novel manner. Researchers have already started to examine possible realizations of these salient features of earthquakes by employing the models exhibiting self-organized criticality [20]. Also, there is an attempt to understand inappropriateness of imposing space-time windows to analyze seismicity based on the idea of the scale-free earthquake network [21].

In this article, we study the structural and dynamical aspects of the earthquake network. Differently from the small-world picture, we address our interest to *directedness* of the network. In particular, we investigate the statistical property of the degrees of separation. Note that this is a highly nontrivial issue, since given a pair of vertices there are many different directed paths linking them. The degree of separation, $l$, taken there should be the minimum number of arcs among such paths. As a typical example, consider a process linking five vertices: $\cdots v_1 \to v_2 \to v_2 \to v_3 \to v_2 \to v_3 \to v_4 \to v_2 \to v_5 \to v_1 \to \cdots$, which has one loop at $v_2$ and one triple arcs between $v_2$ and $v_3$. In the small-world picture, where directedness is ignored, the value of the degrees of separation between $v_1$ and $v_4$ is 2. On the other hand, there are two paths linking $v_1$ and $v_4$: one has 6 links ($v_1 \to v_2 \to v_2 \to v_3 \to v_2 \to v_3 \to v_4$) and the other has 3 ($v_4 \to v_2 \to v_5 \to v_1$). Therefore, the degree of separation between $v_1$ and $v_4$ is 3. (See Fig. 1.) This directed-network picture may allow us to explore the property of the degrees of separation between two arbitrarily chosen earthquakes on the network (i.e., vertices). Here, we report a new discovery that such correlation on the earthquake network is long-ranged and obeys a power law with respect to the degrees of separation.

The procedure for constructing the earthquake network is as follows. The district of southern California, which we investigate here, is divided into cubic cells with size $10 \, \text{km} \times 10 \, \text{km} \times 10 \, \text{km}$. Then the earthquake catalog made available by the Southern California Earthquake Data Center (http://www.scecdc.scec.org/ catalogs.html) is analyzed. In the data, the region covered is 29°15.25'N–38°49.02'N latitude and 113°09.00'W–122°23.55'W longitude with the maximal depth (of the foci of the observed earthquakes) 57.88km. The period is between 00:25:8.58 on January 1, 1984 and 15:23:54.73 on December 31, 2002. With the above-mentioned cell size, the total number of independent vertices is 3076, whereas the total number of arcs is 375886. As shown in Refs. [18,19], the distribution of connectivities, $p(k)$, which corresponds to the number of vertices with $k$ arcs, decays as a power law of the Zipf-Mandelbrot form [15-17,22]

$$p(k) \sim (k + k_0)^{-\gamma}, \qquad (1)$$

where $k_0 \cong 1.70$ and $\gamma \cong 1.33$. There is a phenomenological reason behind this result. The seismic data tell us that aftershocks associated with a mainshock empirically tend to return to the locus of the mainshock geographically. The stronger a mainshock is, the more it is followed by aftershocks, yielding a larger value of connectivities. On the other hand, the Gutenberg-Richter law [23] implies that (cumulative) frequency of earthquakes decays as a power law with respect to the value of moment. This scale-free nature, therefore, leads to the asymptotic power-law distribution in Eq. (1). In the language of complex evolving networks, a mainshock plays a role of a "hub" and preferential attachment [15-17] is realized by aftershocks.

Now, we have reanalyzed the earthquake network thus constructed by taking its directedness into account. We have performed random sampling of 10000 pairs of vertices from the network and have counted the number of pairs, $n(l)$, which are linked by $l$ arcs (including loops) in the directed-network picture. (Due to limitation of our computational power, we have analyzed up to $l = 100$, since the full treatment is a heavy combinatorial problem.) The result is depicted in Figs. 2 and 3. It shows that similarly to Eq. (1) $n(l)$ also obeys the Zipf-Mandelbrot law

$$n(l) \sim (l + l_0)^{-\delta}. \tag{2}$$

This asymptotic power law leads to an important conclusion that "event correlation" on the directed earthquake network is long-ranged and scale-free. In this respect, it is essential to recall the fundamental difference between the distance on the network (the degrees of separation) and the Euclidean distance. The law in Eq. (2) may be interpreted in connection with spatial-temporal criticality/complexity of seismicity. It is an empirical fact recognized by performing data analysis [9] that the Euclidean distance between two successive earthquakes is often very large. If such two events were independent and uncorrelated, no simple laws could be expected to exist. However, on the contrary, the present result indicates that the directed earthquake network in fact follows a definite statistical law.

In conclusion, we have found a new scale-free nature of the directed earthquake network with respect to the degrees of separation. The long-tailed distribution in Figs. 2 and 3 suggests that there exists a yet unknown mechanism in seismicity, which makes "event-event correlation" on the network extremely persistent both spatially and temporally. It may also be of interest to examine if this empirical law can be realized in the models exhibiting self-organized criticality [1,2].

**Acknowledgment**

S. A. was supported in part by the Grant-in-Aid for Scientific Research of Japan Society for the Promotion of Science.

# Figure Captions

Fig. 1    A schematics description of the earthquake network. The degree of separation between $v_1$ and $v_4$ is 3 in the directed-network picture.

Fig. 2    (a) The linear plot of the number of pairs of vertices (out of 1000 randomly sampled pairs) with respect to their degrees of separation in the directed-network picture. The solid line represents the law in Eq. (2) with $l_0 \cong 0.95$ and $\delta \cong 0.67$. All quantities are dimensionless.

(b) The log-log plot of Fig. 2 (a).

Fig. 3    The log-log plot of the *complementary* cumulative number defined by $N(<l) = \int_0^l dl' \, n(l')$ associated with $n(l)$ in Fig. 2. All quantities are dimensionless.